\documentclass[prl,twocolumn,preprintnumbers,amsmath,amssymb,superscriptaddress,floatfix,showpacs]{revtex4}

\usepackage{graphicx}
\usepackage{dcolumn}
\usepackage{bm}
\usepackage{nicefrac}

\begin{document}


\title{Direct observation of $t_{2g}$ orbital ordering in magnetite}

\author{J. Schlappa}
\affiliation{{II.} Physikalisches Institut, Universit{\"a}t zu K{\"o}ln,
Z{\"u}lpicher Str.~77, D-50937 K{\"o}ln, Germany}
\author{C. Sch{\"u}{\ss}ler-Langeheine}%
\affiliation{{II.} Physikalisches Institut, Universit{\"a}t zu K{\"o}ln,
Z{\"u}lpicher Str.~77, D-50937 K{\"o}ln, Germany}
\author{C.~F. Chang}%
\affiliation{{II.} Physikalisches Institut, Universit{\"a}t zu K{\"o}ln,
Z{\"u}lpicher Str.~77, D-50937 K{\"o}ln, Germany}
\author{H. Ott}%
\affiliation{{II.} Physikalisches Institut, Universit{\"a}t zu K{\"o}ln,
Z{\"u}lpicher Str.~77, D-50937 K{\"o}ln, Germany}
\author{A.~Tanaka}
\affiliation{Department of Quantum Matter, ADSM, Hiroshima University,
Higashi-Hiroshima 739-8530, Japan}
\author{Z. Hu}%
\affiliation{{II.} Physikalisches Institut, Universit{\"a}t zu K{\"o}ln,
Z{\"u}lpicher Str.~77, D-50937 K{\"o}ln, Germany}
\author{M. W. Haverkort}%
\affiliation{{II.} Physikalisches Institut, Universit{\"a}t zu K{\"o}ln,
Z{\"u}lpicher Str.~77, D-50937 K{\"o}ln, Germany}
\author{E. Schierle}
\affiliation{Institut f{\"ur} Experimentalphysik, Freie Universit{\"a}t Berlin,
Arnimallee 14, D-14195 Berlin, Germany}
\author{E. Weschke}
\affiliation{Institut f{\"ur} Experimentalphysik, Freie Universit{\"a}t Berlin,
Arnimallee 14, D-14195 Berlin, Germany}
\author{G. Kaindl}
\affiliation{Institut f{\"ur} Experimentalphysik, Freie Universit{\"a}t Berlin,
Arnimallee 14, D-14195 Berlin, Germany}
\author{L. H. Tjeng}%
\affiliation{{II.} Physikalisches Institut, Universit{\"a}t zu K{\"o}ln,
Z{\"u}lpicher Str.~77, D-50937 K{\"o}ln, Germany}

\date{\today}

\begin{abstract}
Using soft-x-ray diffraction at the site-specific resonances in
the Fe $L_{2,3}$ edge, we find clear evidence for orbital and charge
ordering in magnetite below the
Verwey transition. The spectra show directly that the
(00$\nicefrac{1}{2}$) diffraction peak (in cubic notation) is
caused by $t_{2g}$ orbital ordering at octahedral
Fe$^{2+}$ sites and the (001) by a spatial modulation of the $t_{2g}$ occupation.
\end{abstract}

\pacs{71.30.+h,71.45.Lr,61.10.Nz,78.70.Dm}

\maketitle

Magnetite, Fe$_3$O$_4§$, is one of the most fascinating materials
in solid state physics. Besides being the first magnetic material
known to the mankind, see e.g. \cite{matthis:81a}, it is also the
first example for an oxide to show a first order anomaly in the
temperature dependence of the electrical conductivity at $T_V
\approx 120$ K, the famous Verwey transition \cite{verwey:39a}.
It is accompanied by a structural phase transition from the cubic
inverse spinel to a distorted structure, leading to the
appearance of superstructure diffraction peaks mainly
characterized by the wave vectors (001) and (00$\nicefrac{1}{2}$)
(the notation throughout this paper refers to the cubic
high-temperature unit cell with $a$ = 8.396 \AA). One usually
connects this transition with charge ordering of Fe$^{2+}$
and Fe$^{3+}$ ions on the octahedrally coordinated, so called,
B-sites \cite{tsuda:83book}.

Very recently, Wright, Attfield and Radaelli found from their
neutron and synchrotron-x-ray diffraction measurements an
intriguing pattern of shorter and longer bond lengths between
B-site Fe and oxygen ions below $T_V$, which they interpreted in
terms of a particular charge order \cite{wright:01a,wright:02a}.
This result has triggered a flurry of new theoretical and
experimental efforts. Particularly important hereby are the LDA+$U$ band
structure studies by Leonov \textit{et al.} and Jeng \textit{et
al.} who made the conjecture that it is the $t_{2g}$ orbital
occupation and order which are the characteristic order
parameters of the system \cite{leonov:04a,jeng:04a} rather than
the order of the total charge itself as its variation is rather
small. Yet, most of the follow-up experimental investigations
\cite{subias:04a,nazarenko:06a} are still focused on the charge
order issue. In a soft x-ray diffraction study of Huang \textit{et al.} the importance
of the orbital and charge order issue was recognized \cite{huang:06a}, but the experiments at the oxygen $1s \rightarrow 2p$ ($K$) resonance are not specific to the Fe $t_{2g}$ orbital order.

Here we report on our resonant soft-x-ray diffraction (RSXD) study
on Fe$_3$O$_4$ using photons in the vicinity of the Fe $L_{2,3}$
absorption edges. This technique directly probes the
$3d$-electronic states of the transition-metal ion via the dipole
allowed Fe $2p \rightarrow 3d$ ($L_{2,3}$) excitation involved in
the scattering process
\cite{elfimov:99a,castleton:00a,thomas:04a,schuessler:05a}. This
has enabled us to obtain clear evidence for the presence of the Fe
$t_{2g}$ orbital and charge order, thereby providing experimental support for
the ideas brought forward by the LDA+$U$ studies
\cite{leonov:04a,jeng:04a}.

\begin{figure}
\includegraphics[scale=1,clip,bb=109 509 485 757,width=7.5cm]{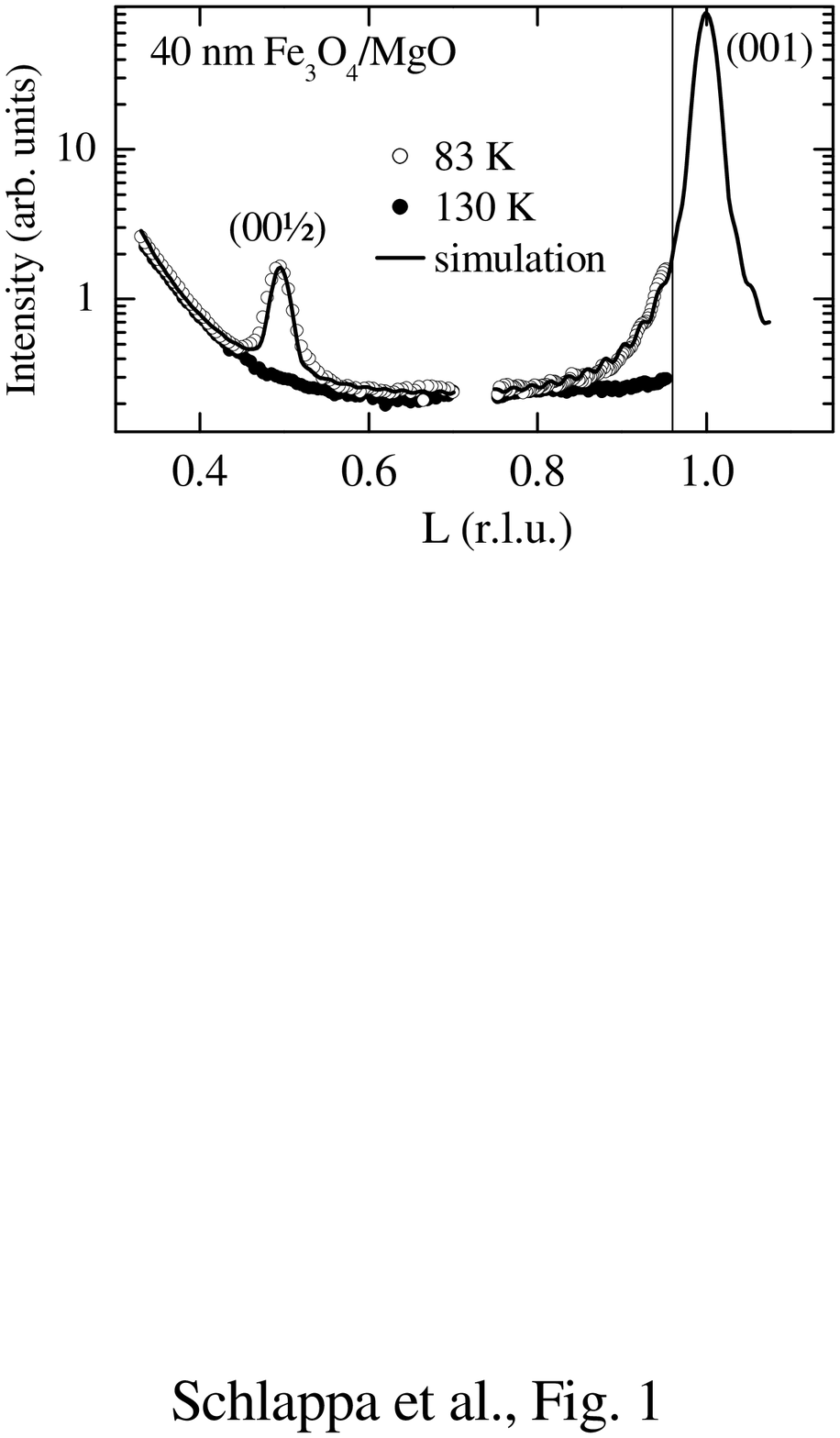}
\caption{Scan along L below (open symbols) and above (filled
symbols) $T_V$ recorded at 708.5 eV photon energy with $\sigma$
polarization. The solid line through the data points is a
simulation. The vertical line at L=0.96 denotes the maximum
possible momentum transfer at this energy.}
\end{figure}

For this study we chose a magnetite film rather than a bulk
crystal for two reasons. Firstly, diffraction features in a thin
film are broadened along the surface normal. This broadening
allows us to study the resonance of the (001) peak, whose peak
maximum cannot be reached at the $L_{2,3}$ resonance. Secondly,
the volume probed by the scattering experiment is determined by
the film thickness rather than by the photon penetration depth.
Since the latter changes strongly across resonances in the soft
x-ray range, this change may obscure the resonance effect. A
40-nm film of Fe$_3$O$_4$ was grown ex-situ by molecular-beam
epitaxy (MBE) on epi-polished MgO. In order to separate the
specular reflectivity from the $c$-direction of Fe$_3$O$_4$, a
substrate with about 6$^\circ$ miscut was used. The film was
characterized by resistivity measurements and shows a sharp
Verwey transition at 115 K. RSXD experiments were carried out at
the BESSY beamline UE52-SGM using the UHV diffractometer built at
the Freie Universit\"at Berlin. The sample was oriented such that
two of the cubic axes of the room temperature structure were
parallel to the diffraction plane. The incoming light
polarization was either perpendicular ($\sigma$-polarization) or
parallel ($\pi$-polarization) to the diffraction plane.

Fig.~1 shows a $q$-scan along the [001] or L direction of the
reciprocal space taken at 83 K (open symbols), i.e. below $T_V$,
with 708.5 eV photons, revealing the (00$\nicefrac{1}{2}$)
diffraction peak, and the onset of the (001). Even though the
maximum of the (001) cannot be reached at the Fe-$L_{2,3}$
resonance, where the maximum momentum transfer is L=0.96 as
indicated by the vertical line in Fig.~1, the broadening of the (001) peak transfers some
intensity into the reachable momentum space. The solid line
through the data points is a simulation including two peaks of
the same shape but different intensity and the Fresnel reflectivity, which causes the
increasing background at low L-values. This model describes the
data reasonably well, even though the oscillations of the Laue
profile are not resolved in the data. The intensity of the (001)
peak is about ten times that of (00$\nicefrac{1}{2}$). Both peaks
disappear when the sample is heated to 130 K (filled symbols),
i.e. above $T_V$. From the peak widths we determine the
correlation length perpendicular to the surface to be 37 nm,
which is essentially the full film thickness. The in-plane
correlation length of 10 nm is most probably limited by the
formation of crystalline domains \cite{eerenstein:02b}. The
influence of the stepped substrate on the structure of the film
seems to be weak, since the in-plane correlation length is about
2.4 times larger than the MgO terrace width.

In order to identify the origin and nature of the two diffraction
peaks, we recorded their intensity as a function of photon
energy. The background below the (00$\nicefrac{1}{2}$) peak was
interpolated from the resonance behavior at L = 0.35, 0.4, 0.6,
and 0.7. The spectrum for the (001) peak was recorded at L = 0.95
with the background intensity taken at L = 0.8.

\begin{figure}
\includegraphics[scale=1,clip,bb=82 432 524 735,width=8cm]{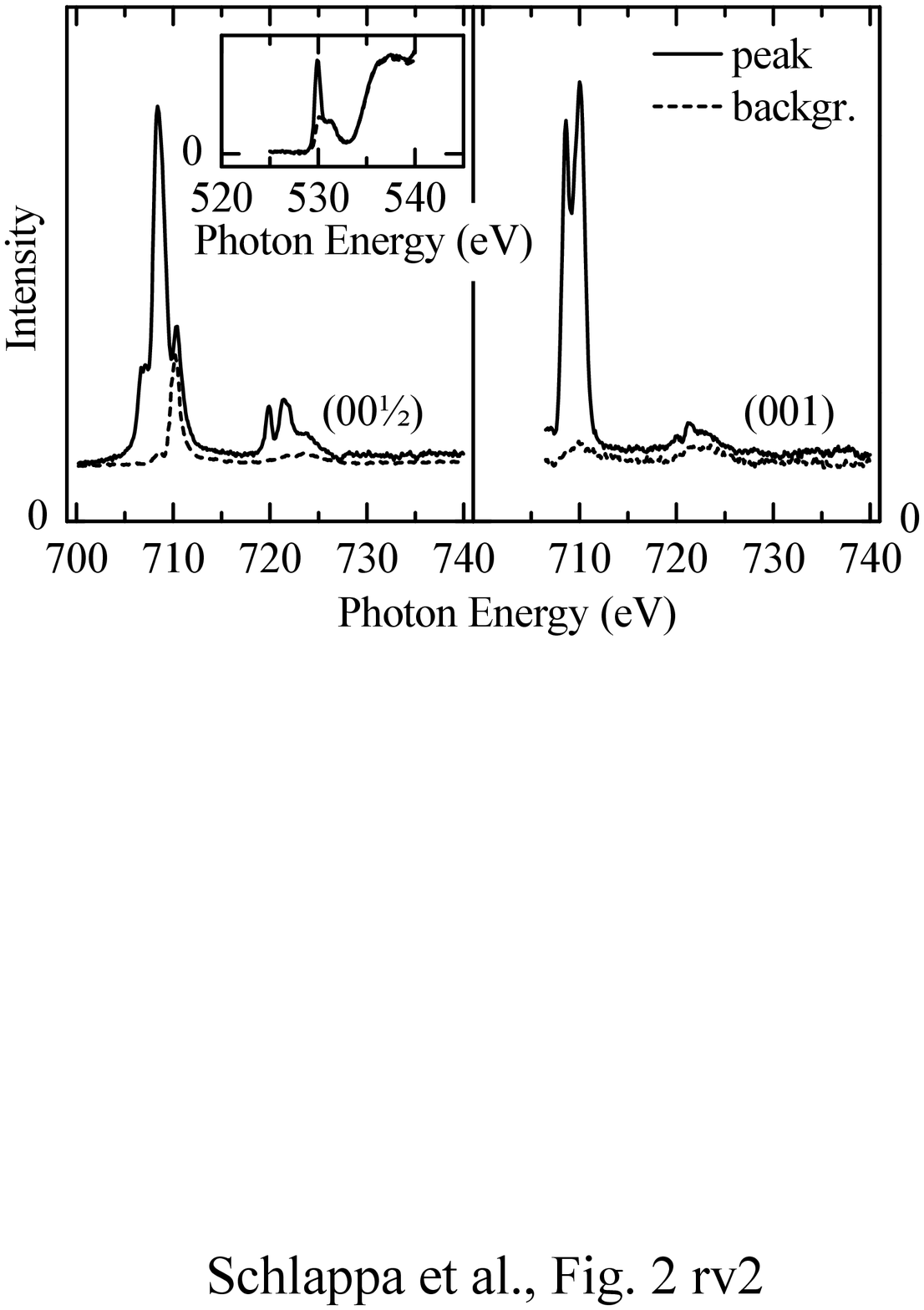}
\caption{Spectra of the (00$\nicefrac{1}{2}$) (left) and (001)
(right) diffraction peaks (solid lines) and the corresponding
backgrounds (dashed lines) recorded with $\pi$ polarized light.
The inset shows the (00$\nicefrac{1}{2}$) spectra recorded at the
oxygen $K$-edge for comparison. All spectra are plotted on the
same vertical scale.}
\end{figure}

The spectra obtained from the peaks and backgrounds are presented
in Fig. 2. The shapes of the spectra for the two peaks are
distinctly different: the $L_3$ part of the (00$\nicefrac{1}{2}$)
spectrum has a sharp resonance maximum at 708.4 eV, while the
spectrum of the (001) peak shows a double peak structure, one
peak at 708.6 eV, i.~e. almost the same energy as the maximum of
the (00$\nicefrac{1}{2}$) spectrum, and the second one at 710.1
eV. The $L_2$ parts of the spectra, on the other hand, are more
similar with two maxima at 720 eV and 721.4 eV. The inset shows
the corresponding data from the oxygen-$K$ edge as a reference.

In order to determine to which extent our results are influenced
by film properties, we have carried out experiments using a film
twice as thick as the one presented above and grown on a flat MgO
substrate. We have seen that the reflectivity background below
the (00$\nicefrac{1}{2}$) peak was larger and that the onset of
the much narrower (001) peak was less well defined. These
findings however, justify precisely our motivation to use thin
films on a miscut substrate rather than a bulk crystal with a low
index surface as outlined in the experimental paragraph above.
The key data of this thicker film are identical to those of the
40 nm film: both the (00$\nicefrac{1}{2}$) and (001) diffraction
peaks can be observed below $T_V$ and both disappear above $T_V$;
the energy dependence of their intensities are very similar to
the ones shown in Fig.~2. This indicates that the results
presented are not a particularity of a magnetite film, but that
they are indeed representative for the bulk material.

\begin{figure}
\includegraphics[scale=1,clip,bb=60 251 511 750,width=8cm]{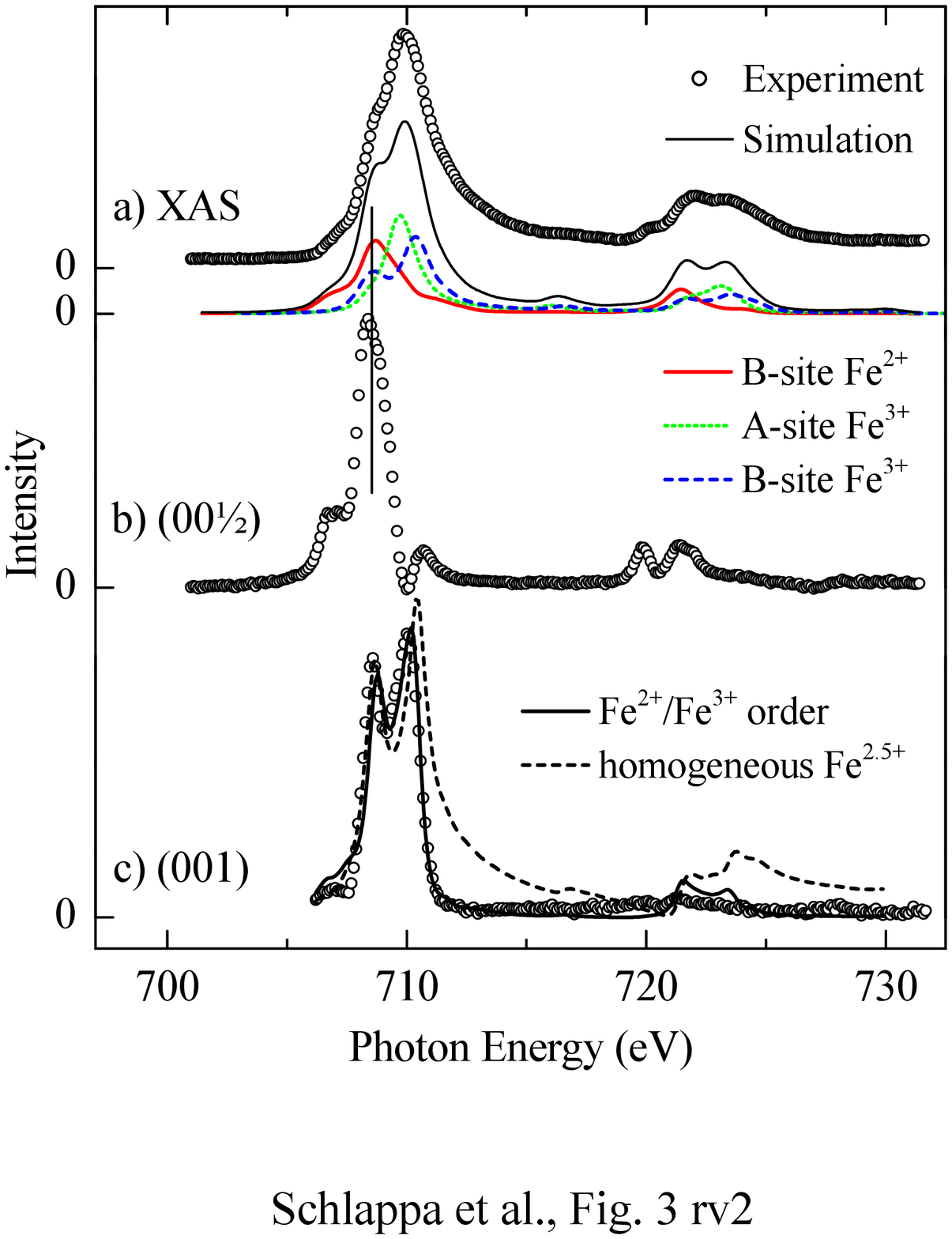}
\caption{(color online). a) Experimental Fe $L_{2,3}$ XAS spectra
of Fe$_3$O$_4$ (symbols). The black thin solid line is a
simulation, which is constructed as the sum of the spectra from
the B-site Fe$^{2+}$ (red thick solid line), A-site Fe$^{3+}$
(green dotted line), and B-site Fe$^{3+}$ (blue dashed line)
ions; b) energy dependence of the (00$\nicefrac{1}{2}$)
diffraction peak intensity; c) energy dependence of the (001)
intensity (symbols), the solid line is a simulations for a
charge-order scenario and the dashed line for a homogeneously
mixed-valent scenario, both scaled to the height of the first resonance peak.}
\end{figure}

In Fig.~3 we present the resonance spectra of the two diffraction
features after background subtraction together with the Fe
$L_{2,3}$ XAS spectrum. The spectral lineshapes for the two
diffraction peaks are clearly different from the XAS spectrum
[symbols in Fig. 3a)]. To interpret the spectroscopic results, we
resort to the existing literature on the XAS of Fe$_3$O$_4$
\cite{park:thesis,kuiper:97a,chen:04a} with emphasis on the
strong magnetic circular and linear dichroism
\cite{kuiper:97a,chen:04a} effects therein. The sharp structures
in the dichroic spectra allow for a clear decomposition of the
XAS spectrum in terms of contributions coming from the different
Fe sites in magnetite. This decomposition is given in Fig.~3a):
the spectrum from the B-site Fe$^{2+}$ ion is shown by the red
thick solid line, the A-site Fe$^{3+}$ by the green dotted line,
and the B-site Fe$^{3+}$ by the blue dashed line.  The maximum in
the $L_3$ white line of each of these ions occurs at quite
different energies, an aspect which we will utilize next to
interpret the spectra (001) and (00$\nicefrac{1}{2}$) diffraction
peaks. 

The maximum of the (00$\nicefrac{1}{2}$) diffraction peak
spectrum occurs at essentially the same energy as that of the
B-site Fe$^{2+}$ XAS. This means that the (00$\nicefrac{1}{2}$)
diffraction peak is due to an order, which involves \emph{only}
B-site Fe$^{2+}$ ions. This in turn implies that orbital order of
the $t_{2g}$ electrons is at play here, since this is the only
degree of freedom available which could make one B-site Fe$^{2+}$
ion to be different from another B-site Fe$^{2+}$ ion. In $P2/c$
or $Cc$ symmetry the (00$\nicefrac{1}{2}$) diffraction peak is
glide-plane forbidden with the consequence that only off-diagonal
elements of the scattering tensors contribute. These elements can
be expected to be much stronger for the Fe$^{2+}$ ($d^6$) ions
than for the more spherical Fe$^{3+}$ ($d^5$) ions, explaining
indeed why the resonant enhancement of the (00$\nicefrac{1}{2}$)
peak occurs only for photon energies which are characteristic for
the Fe$^{2+}$ sites. Our result that the resonant
enhancement of the (00$\nicefrac{1}{2}$) peak occurs only for
photon energies which are characteristic for the Fe$^{2+}$ sites
thus provides direct experimental proof for orbital-order as
predicted in Refs.~\onlinecite{leonov:04a,jeng:04a}.  We note that this orbital-order finding does
not exclude the existence of charge-order associated with other,
not glide-plane forbidden peaks related to the doubling of the
unit cell along $c$ as observed in Ref.~\onlinecite{nazarenko:06a}.

In contrast to (00$\nicefrac{1}{2}$), the (001) spectrum shows
two peaks which essentially occur at the energies of the maxima
of the two B-site resonances. Such a double-peak structure is
exactly the resonance shape to be expected from charge order
involving the two B-site ions. The experiment is sensitive to
differences in the scattering amplitudes of the different ions.
For two resonances well separated in energy, as it is the case
for the B-site Fe$^{2+}$ and Fe$^{3+}$ ions, this difference is
large near the two resonance maxima leading to such a two-peak
feature. To illustrate this we show as the solid line in Fig.~3c)
the result of a simulation using the complex scattering
amplitudes for the B-site ions, which we have extracted from the
subspectra in Fig.~3a). We find a very good agreement between
experiment and simulation.

We would like to remark that the (001) is a structurally allowed
Bragg peak, so that it should be visible even in the case of a
homogeneously mixed-valent system, i.e. if there were no charge
order. Since the scattering amplitude of the Fe sites changes
across resonance, even a pure structural peak will show a
resonance effect. In the structure proposed by Wright \textit{et al.} the B-site contribution to the amplitude for (001) is $ \approx 4.1 f_{2+}(\omega) -3.7 f_{3+}(\omega)$ \cite{wright:02a}
with the $f$s describing the scattering amplitudes for the
different ions. In a charge-ordered case $f_{2+}(\omega) \neq
f_{3+}(\omega)$ and for energies where one resonance is
dominating the charge order contribution to the observed
intensity would be of the order of $16 |f|^2$. The structural
contribution is the one that would remain even if all sites were
identical. It gives an intensity of the order of $(4.1 - 3.7)^2 |f|^2$,
i.e. about 100 times weaker than the intensity expected in the case of charge ordering.

Since (001) is stronger than (00$\nicefrac{1}{2}$), the charge
order scenario appears to be more plausible, but a quantitative
comparison requires to know the absolute sizes of the scattering
tensor elements. The identification of (001), however, is
possible even without intensity reference from its 
spectral line shape. The resonant scattering amplitudes consist
of a complex energy-dependent part $f'(\omega)+i f''(\omega)$
plus an energy-independent part $f_0$. The scattering amplitude for the charge order scenario is dominated by the difference between $f_{2+}$ and $f_{3+}$. In this difference the $f_0$ term, which is
essentially the same for both iron valences, cancels out. It does, however,
contribute to the structurally scattered intensity. The latter is $|f_0+f'+if''|^2$ and thus contains
a mixed term $2 f_0 f'$. Since $f'$ is asymmetric going through a
minimum at the low-energy side of the absorption maximum, the
mixed term is asymmetric too; its relative contribution to the
total intensity depends on the ratio between $f_0$ and $f'$,
$f''$. We estimated this ratio by scaling the edge jump in the
XAS data to the tabulated values for $f''$ \cite{henke:95a}. A
simulation of the spectral shape for a no-charge-order scenario including the mixed-term contribution is presented as the dashed line in Fig. 3c). The B-site part of the absorption spectrum of this hypothetical homogeneous system was taken as the average of the two B-site spectra in Fig. 3a). For better
comparison the two simulated spectra were scaled to the same intensity at the low-energy peak. The asymmetry in the simulated no-charge-order spectrum is considerable
and clearly not observed in the experimental data. We therefore
conclude that (001) is dominated by B-site charge order.

As indicated by the LDA+$U$ band structure calculations, the charge modulation is rather in the $t_{2g}$ count and not so much in the total $d$-electron count on the Fe sites. The reason is that modulation of the $t_{2g}$ occupation is partially screened by charge transfer from oxygen neighbors to the empty $e_g$ states \cite{leonov:04a}. For the $t_{2g}$ states LDA+$U$ finds that the modulation amounts to 0.7 electron, i.e. almost a full electron, while the total $3d$ modulation is only 0.23 electrons. As a check of our analysis, in which we have used the cluster model calculations \cite{tanaka:94a} to simulate the XAS spectra in Fig.~3, we have looked at the $t_{2g}$ occupations in the FeO$_6$ clusters constructed for the Fe$^{2+}$ and Fe$^{3+}$ ions: we find a $t_{2g}$ occupation difference of 0.86 electron. This is also close to unity, i.e. very consistent with the LDA+$U$ results. In fact, one must expect the cluster calculation to slightly overestimate the occupation difference since inter-cluster hopping or band-formation effects are not included.

Summarizing, using resonant soft x-ray diffraction at the Fe-$L_{2,3}$ threshold from a film of magnetite grown on stepped MgO we were able to establish directly that there is $t_{2g}$ orbital order on the B-sites Fe$^{2+}$ ions as predicted by LDA+$U$ calculations, and that this is reflected in the existence of the (00$\nicefrac{1}{2}$) diffraction peak. We also showed that charge order of the B-site ions is the dominant reason for the appearance of the (001) diffraction peak, thereby settling the contradicting claims from Fe $K$-edge resonant diffraction studies.

We are grateful to M. Braden, H.-H. Hung, P. G. Radaelli, P.
Abbamonte, G. A. Sawatzky, and especially D. I. Khomskii for very
helpful discussions and L. Hamdan for her skillful technical
assistance. The research in Cologne is supported by the Deutsche
Forschungsgemeinschaft through SFB 608.


\end{document}